\documentclass[fleqn,usenatbib]{mnras}

\usepackage{newtxtext,newtxmath}

\usepackage[T1]{fontenc}

\DeclareRobustCommand{\VAN}[3]{#2}
\let\VANthebibliography\thebibliography
\def\thebibliography{\DeclareRobustCommand{\VAN}[3]{##3}\VANthebibliography}

\usepackage{graphicx}	
\usepackage{amsmath}	

\title[Tidally induced double bars]{A new tidal scenario for double bar formation}

\author[Semczuk et al.]{%
Marcin Semczuk,$^{\!\!1,2,3}$
Ewa L. \L{}okas,$^{\!\!4}$ 
Adriana de Lorenzo-Cáceres, $^{\!\!5,6}$
and
E. Athanassoula $^{\!\!7}$
\smallskip
\\
$^1$ Departament de Física Qu\`antica i Astrof\'isica (FQA), Universitat de Barcelona (UB),  c. Mart\'i i Franqu\`es, 1, 08028 Barcelona, Spain \\
$^2$ Institut de Ci\`encies del Cosmos (ICCUB), Universitat de Barcelona (UB), c. Mart\'i i Franqu\`es, 1, 08028 Barcelona, Spain \\
$^3$ Institut d'Estudis Espacials de Catalunya (IEEC), c. Gran Capit\`a, 2-4, 08034 Barcelona, Spain\\
$^4$ Nicolaus Copernicus Astronomical Center, Polish Academy of Sciences, Bartycka 18,00-716 Warsaw, Poland \\
$^5$ Instituto de Astrofísica de Canarias, Calle Vía Láctea s/n, 38205 La Laguna, Tenerife, Spain \\
$^6$ Departamento de Astrofísica, Universidad de La Laguna, 38200 La Laguna, Tenerife, Spain \\
$^7$ Aix Marseille Université, CNRS, CNES, LAM, Marseille, France
}
\date{submitted to MNRAS Letters}

\pubyear{2023}

\begin{document}
\label{firstpage}
\pagerange{\pageref{firstpage}--\pageref{lastpage}}
\maketitle

\begin{abstract}
Double bars make up a significant fraction of barred galaxies. We propose a new formation scenario for double bars that involves tidal interactions. We demonstrate the viability of this scenario using two examples of simulated galaxies from run TNG50-1 of the IllustrisTNG project. In the proposed scenario the inner bar forms first, either in isolation, via instabilities, or through previous tides. The outer bar forms later from the material that is tidally distorted by a strong interaction. The inner and outer bars formed this way rotate with different pattern speeds and can be mistaken for a single bar when their phases align. The double-barred structure is stable and can last for at least 3 Gyr. The inner bars of the tidally induced double bars can also have big sizes, which can possibly explain the origin of sizable inner bars recently found in some galaxies.

\end{abstract}

\begin{keywords}
galaxies: interactions -- galaxies: kinematics and dynamics -- galaxies: spiral -- galaxies: structure -- galaxies: evolution
\end{keywords}



\section{Introduction}
A majority of 60-70 per cent of spiral galaxies in the Local Universe host elongated structures in the form of bars \citep{EskridgeEtal2000, MenendezDelmestreEtAl2007, ShethEtAl2008, Masters2011, Cheung2013, Erwin2018}. Around 30 per cent of barred galaxies are double-barred, i.e. have an inner bar embedded in a larger outer structure \citep{Laine2002, Erwin2004}. The formation and evolution of bars in single-barred galaxies have been well-studied from a theoretical perspective. They can form either in isolation via instabilities (for a review, see \citealt{Athanassoula2013}) or through tidal interactions (e.g. \citealt{Lokas2016}). Some efforts, but fewer have also been undertaken to understand the formation and dynamics of the double bars.

\cite{Maciejewski2000} showed that {\it loop} orbits exist in potentials with two independently rotating bars and can be building blocks of long-lived double-barred structures. It was shown via $N$-body simulations that idealised setups with a rotating bulge \citep{Debattista2007} or a rotating cold disc \citep{Du2015} coexisting with a separate hot disc can produce long-lived double-barred structures. Models engineered in that manner resemble well the properties of observed double-barred systems \citep{Du2016}, however, they lack gas hydrodynamics and do not explain how a galaxy got into a state (that is similar to the initial conditions) that later results in a double-bar formation. \cite{Saha2013} demonstrated that the formation of double bars can occur without the need for a rotating second component, but in single hot discs dominated by a dark matter halo.

\cite{Wozniak2015} showed using $N$-body/hydrodynamical simulations that double bars can be formed in a nuclear stellar disc embedded in an older outer bar. In this scenario, the outer bar fuels gas into the center, where star formation creates the nuclear disc, which forms its own inner bar. A strong prediction of this model is that the mean stellar ages of the inner bar should be smaller than those of the outer one.  

In this letter, we propose a new scenario of the double bar formation, in which the inner bar is created first and tidal interactions induce the outer bar later on. We describe it by discussing two example galaxies from the run TNG50-1 of the magnetohydrodynamical cosmological simulations IllustrisTNG  \citep{Pillepich2019, Nelson2019b, Nelson2019} that have undergone such an evolution. This paper briefly describes the formation and properties of the double bars created in this way, including their lengths, kinematics, and stellar ages.

\section{Formation of the double-barred structure}

\begin{figure*}
	\includegraphics[width=17.5cm]{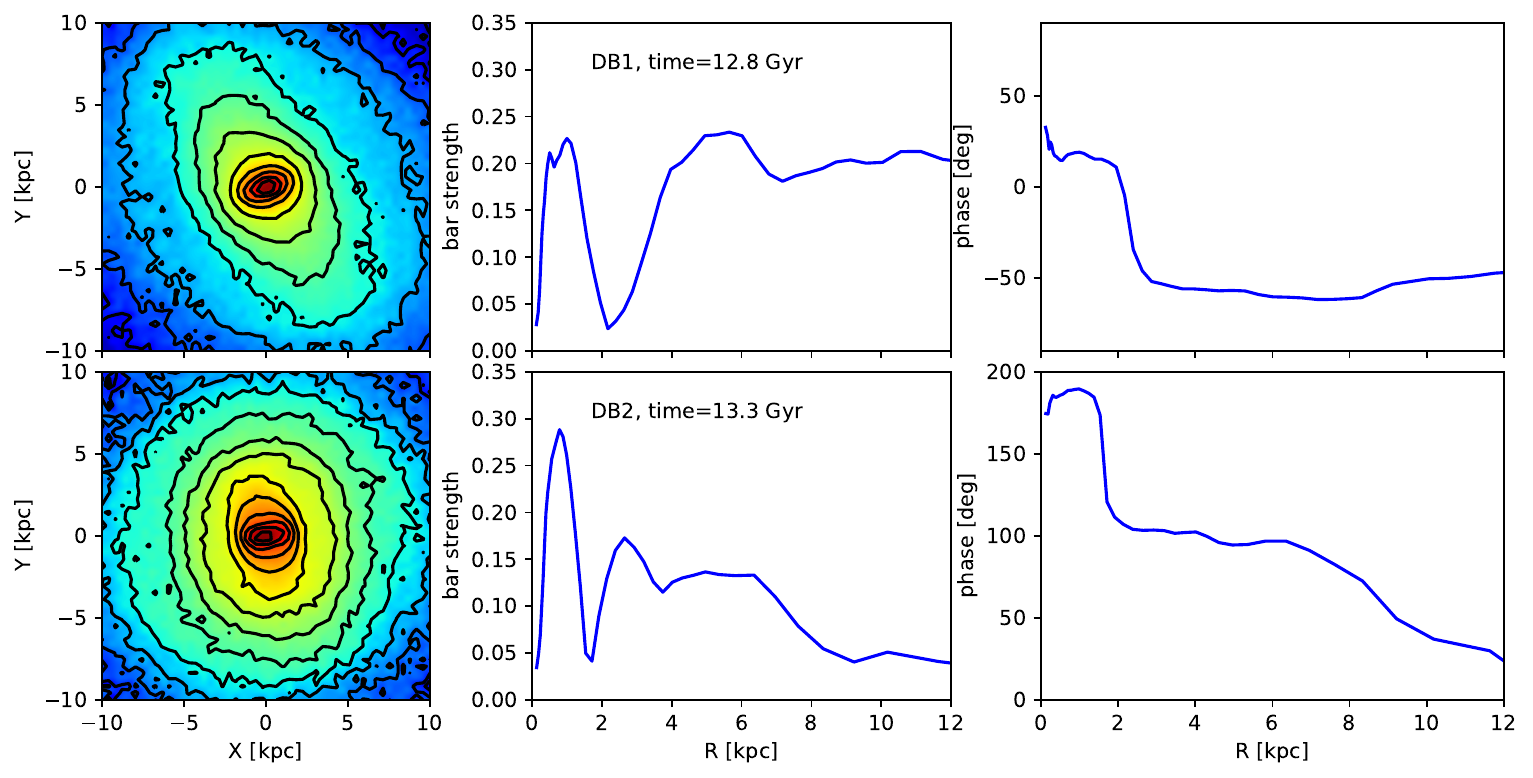}
    \caption{Face-on surface density distributions (left), bar strength (middle), and phase (right) profiles of the two double-barred galaxies from TNG50. For reference, the disc scalenegths of the two galaxies were 4.12 kpc for DB1 and 2.0 kpc for DB2. In addition to the visual impression from the surface density distributions, double bars are visible as two peaks in the bar strength profile and two regions of constant phases in the phase profile.}
    \label{double_bars}
\end{figure*}

\begin{figure*}
	\includegraphics[width=17.5cm]{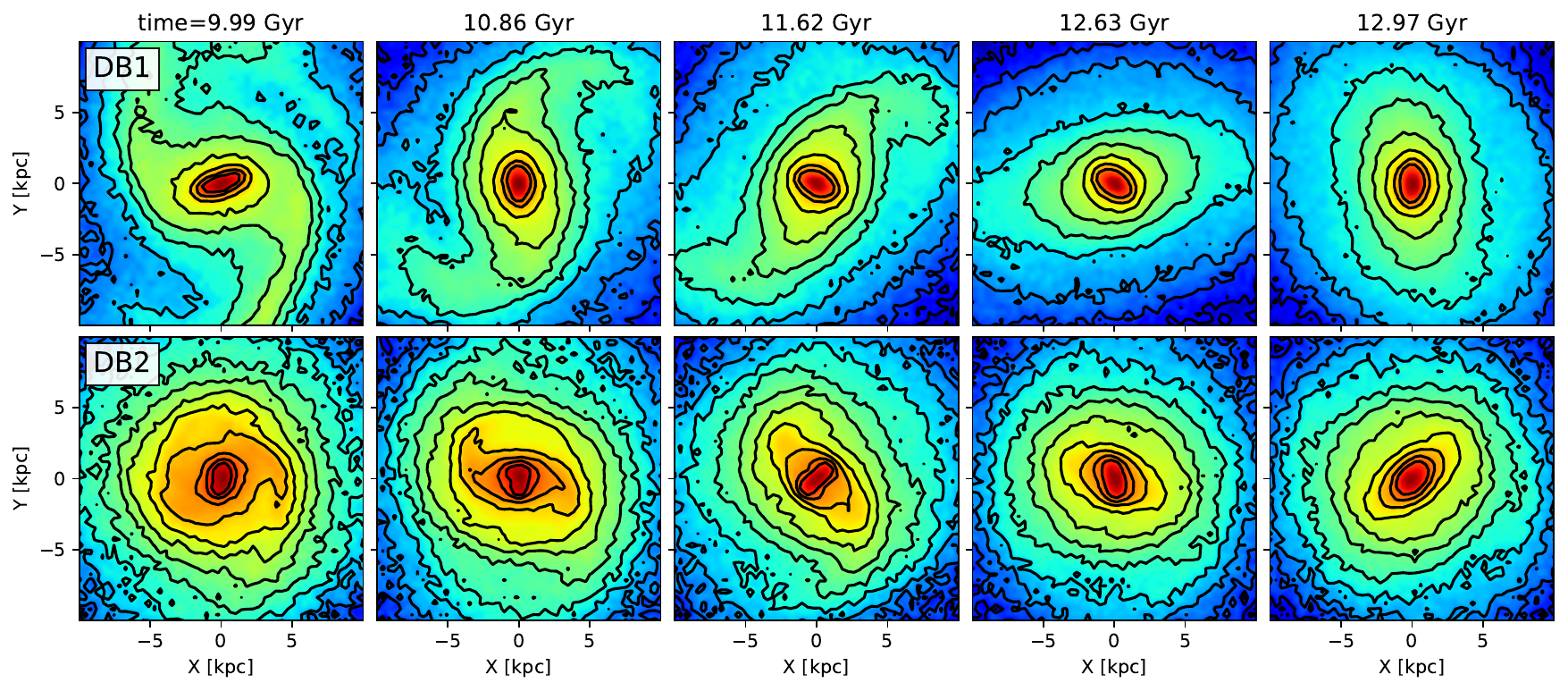}
    \caption{Face-on surface density distribution, for five snapshots of both our double-barred galaxies, showing their evolution from single-barred to double-barred. The three right panels show different relative orientations between the inner and outer bars. When they align, like in the rightmost panel for DB2, the galaxy may appear as single-barred.}
    \label{evolution}
\end{figure*}

\begin{figure}
	\includegraphics[width=\columnwidth]{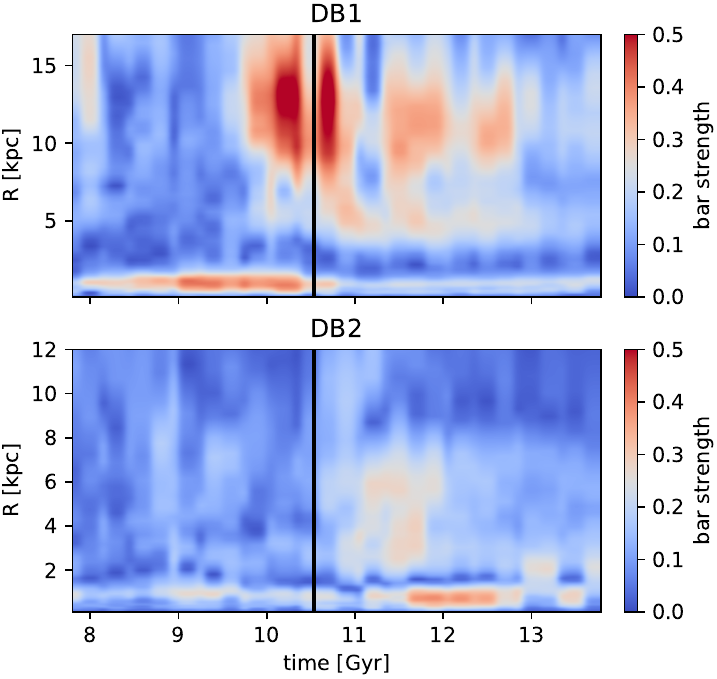}
    \caption{Time evolution of the radial profile of $A_2$ for both double-barred galaxies. Vertical lines mark the time of the pericentre passage of the perturber that induced the outer bar. In both cases, the inner bar was present before this event.}
    \label{A2_panel}
\end{figure}

The example galaxies with double bars formed through the tidal scenario proposed in this paper have IDs at $z=0$ of 294868 and 580406 (hereafter referred to as DB1 and DB2). Both of them were found using the catalogue of barred galaxies from TNG50 compiled by \cite{RosasGuevara2022}. The left panels of Fig.~\ref{double_bars} show their face-on stellar surface density distributions and their radial dependence of the $m=2$ Fourier amplitude and phase. The Fourier analysis was performed using the code of \cite{Dehnen2023} with adaptive radial binning (for details see appendix B of \citealt{Dehnen2023}).
The double-barred morphology is clear in the surface density distributions, as well as in the Fourier $m=2$ analysis. The $m=2$ stellar density amplitude, which is a bar strength measure, shows two regions of higher values and the bar phase has two corresponding regions of constant values.

The formation and evolution of the double-barred structure of these two galaxies can be viewed in Fig.~\ref{evolution} in snapshots of their surface density distribution and in Fig.~\ref{A2_panel} in the time evolution of the bar strength radial profiles. This shows that, before the epoch of the double-bar formation, both galaxies were first single-barred. The galaxy DB1 most likely formed its bar through disc instability, and DB2 through the previous interactions (with a galaxy of ID=373135 at snapshot 50 that passes at 10 kpc near DB2 at 6 Gyr and merges with it at 7 Gyr). After the inner bars of the two galaxies were formed, they both experienced a strong tidal interaction at around 10.5 Gyr. DB1 passed near (<130 kpc) a heavy central galaxy (with a total mass exceeding $10^{13}\;\mathrm{M}_{\odot}$) of a cluster, while DB2 was perturbed by a flying-by galaxy of a mass of approximately $6.3\times10^{10}\;\mathrm{M}_{\odot}$ with a pericenter of about 25 kpc. The effect of these interactions can be seen in Fig.~\ref{A2_panel} as redder regions at higher radii that correspond to tidal spirals, which later weaken, and the remaining oval becomes the outer bar. In the later panels of Fig.~\ref{evolution} we can see that both bars rotate with different pattern speeds, which can be noticed by the change in their relative orientation. When the two bars are aligned, as in the rightmost panel of DB2, such a galaxy can be easily taken for a single-barred one. The change of the relative orientation between the inner and outer bars can also be nicely seen in Fig.~\ref{A2_panel} in the form of the chessboard-like pattern for DB2 at later times. A similar pattern, but with less clarity, can be noticed in Fig. 3 of \cite{Saha2013} in the last few snapshots of their simulations.

\section{Kinematics and stellar ages of the double bars}
\subsection{Kinematics}

\begin{figure}
	\includegraphics[width=\columnwidth]{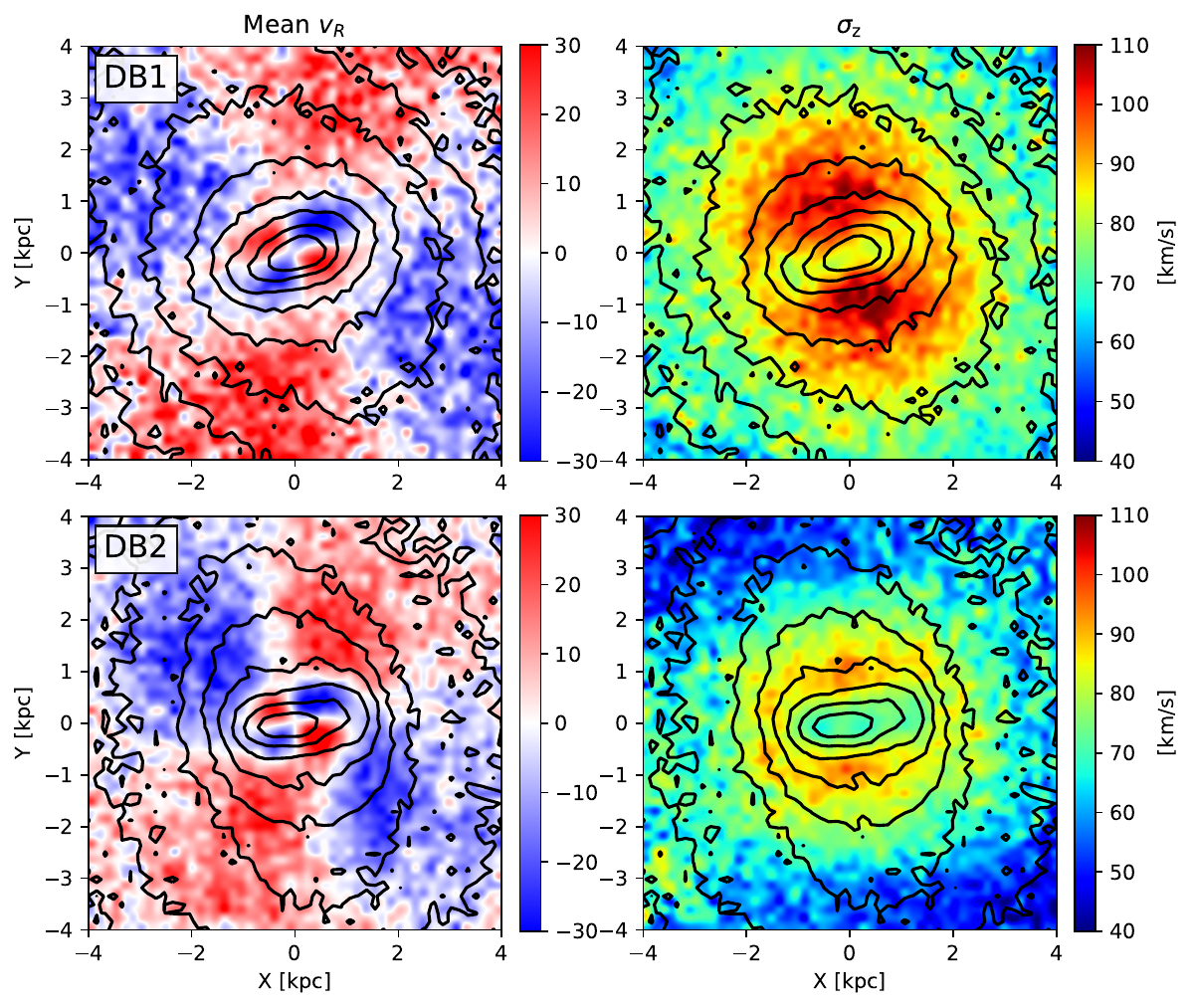}
    \caption{Mean radial velocity distribution (left) and vertical velocity dispersion distribution (right) for simulated double-barred galaxies DB1 at time=12.8 Gyr (top) and DB2 at time=13.3 Gyr (bottom). Overlaid contours mark surface density distribution (as in left panels of Fig.~\ref{double_bars}). Both galaxies show characteristic kinematical features of double bars: the double quadrupole for mean $v_R$ and humps at the minor axis of the inner bar for $\sigma_z$.}
    \label{kinematics}
\end{figure}

One of the most significant kinematical signatures of single galactic bars is a quadrupole structure in cylindrical velocity $v_R$, also called a spider diagram. It originates from the elongated orbits going around the bar shape and turning around at its ends. In the left panels of Fig.~\ref{kinematics} we show the mean $v_R$ distributions for DB1 and DB2 for the same snapshots as shown in Fig.~\ref{double_bars} zoomed-in on the inner 4 kpc. We find that the inner and outer bar have their own independent quadrupole signals, whose boundaries clearly align with the overplotted density contours. These decoupled features are consistent with those found in previous simulations of isolated double bars \citep{Du2016} formed through a different channel.    

\cite{dLC2008} found in observed galaxies that double bars can affect the velocity dispersion field in the form of $\sigma$-hollows associated with the inner bar. \cite{Du2016} later showed using $N$-body simulations of double bars how the observed $\sigma$-hollows can arise from cylindrical $\sigma_z$ humps that are aligned with the minor axis on the inner bar. The origin of $\sigma_z$ was further explored by \cite{Du2017}. In the right panels of Fig.~\ref{kinematics}, 
we show the $\sigma_z$ distribution for DB1 and DB2 with humps that lie on the minor axis of the inner bar. We find that they rotate together with the inner bar. We also find that distributions of $\sigma_R$ and $\sigma_\phi$ (not shown) look very similar to distributions presented by \cite{Du2016} for double-barred simulated galaxies.

\subsection{Stellar ages}

\begin{figure}
	\includegraphics[width=\columnwidth]{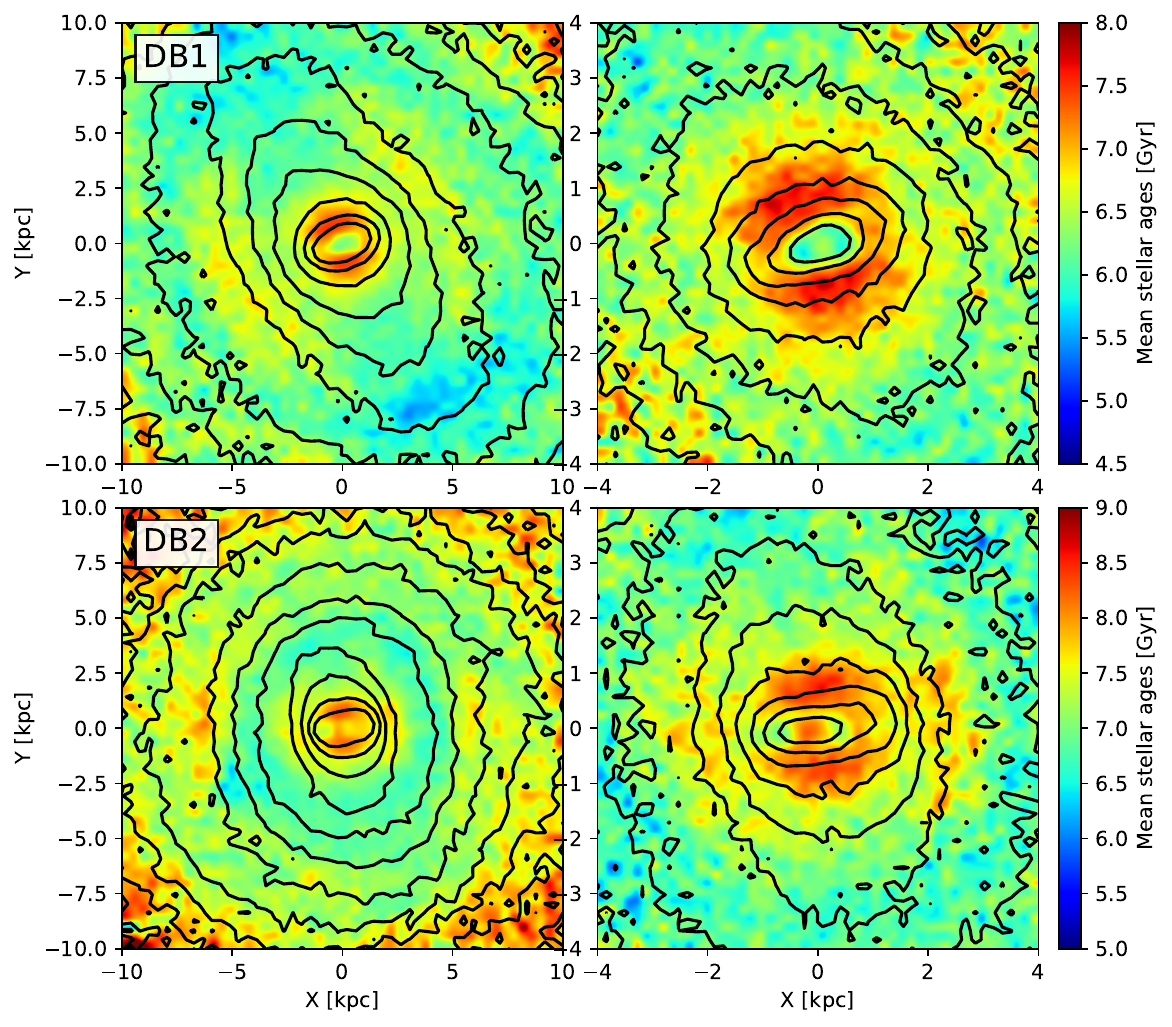}
    \caption{Distributions of the mean stellar ages for simulated double-barred galaxies DB1 at time=12.8 Gyr (top) and DB2 at time=13.3 Gyr (bottom). Overlaid contours mark surface density distribution (as in Fig.~\ref{double_bars}). The right panel shows the zoomed inner 4 kpc region of the left panel, focused on the inner bar. 
    }
    \label{ages}
\end{figure}

In the scenario proposed in this letter, the inner bar is dynamically older than the outer one, i.e. it forms first. This is contrary to the nuclear stellar disc formation scenario, where the outer bar forms first and later funnels gas inward. This forms the nuclear stellar disc, which
later hosts the inner bar. The dynamical age of the bar, however, is not directly linked to the ages of stars. The simulations of \cite{Wozniak2015} predict that in the nuclear stellar disc scenario, the mean ages of stars should be younger, because star formation is an essential ingredient in forming the inner bar. However, as discussed by \cite{dLC2019} in the case of NGC 1291, star formation can still happen outside the inner bar, after this was formed, and therefore ages of the inner bar can be still older than its surroundings. 

In Fig.~\ref{ages} we check what are the mean stellar age distributions of the two simulated tidally induced double bars. We find that the tidal scenario can produce at least two cases of age distributions with respect to the inner bar. For DB1 the stars inside the inner bar are on average younger than outside of it. This is because, after the double-barred structure was created, there was intense enough star formation in the inner bar. At the edge of the inner bar, there is a ring of older stars, and in the outer bar, there is an azimuthal dependence of the mean age. For DB2, the stars inside the inner bar are on average older than those outside of it. Since the tidal scenario of the double bar formation can produce different distributions of stellar ages, such observables cannot be used to reliably discriminate between different ways of forming double bars. On the other hand, this also means that the proposed scenario is able to reproduce a variety of stellar ages found in observed double bars.

\section{Discussion and summary}

\begin{figure}
	\includegraphics[width=\columnwidth]{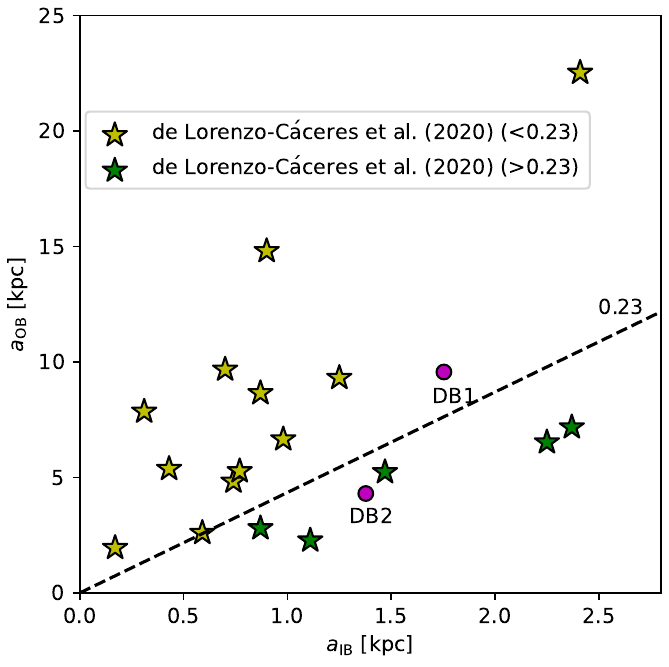}
    \caption{Inner ($a_{\mathrm{IB}}$) and outer ($a_{\mathrm{OB}}$) bar lengths for the simulated galaxies from this paper (DB1 and DB2, magenta points) compared with bar semi-major axes of the observed galaxies from \protect\cite{dLC2020b}(stars). Colours of the data from \protect\cite{dLC2020b} indicate whether the ratio between the inner and outer bar length is $a_{\mathrm{IB}}/a_{\mathrm{OB}}>0.23$ (green) or $a_{\mathrm{IB}}/a_{\mathrm{OB}}<0.23$ (yellow) as defined in that paper.}     
    \label{Lbars}
\end{figure}

Recently \cite{dLC2020b} carefully analysed a sample of 17 double-barred galaxies and found that their sample can be divided into two populations based on the ratio between the inner bar $a_{\mathrm{IB}}$ and the outer bar length $a_{\mathrm{OB}}$. Inner bars in the long inner bars group were found to be larger than some of the outer bars in other galaxies. To check where the simulated tidally induced double bars lie in the plane of $a_{\mathrm{IB}}$ vs $a_{\mathrm{OB}}$ we estimated the bar sizes from the snapshots presented in Fig.~\ref{double_bars} in a following manner. First, we find the maxima of the bar strength corresponding to the inner and outer bars. Then we extend the bar region inwards and outwards and measure the maximal difference of phases in that radial region. We extend this bar region as long as the difference does not exceed 10 degrees or the bar strength reaches the threshold of 0.07. It is similar to the method of finding the bar region in \cite{Dehnen2023}. Although this is a rather rough estimate, comparing the obtained values with the visual impression from Fig.~\ref{double_bars} we find reasonably good agreement (possibly thanks to the lack of spiral arms that could affect such measurements at other times).     

Fig.~\ref{Lbars} compares the estimates described above for the simulated double bars with the results of \cite{dLC2020b}. We note that the estimates of \cite{dLC2020b} done by the photometric decomposition may be slightly higher with respect to estimates done in a similar way to those applied here to the simulations. Both simulated galaxies lie close to the border of $0.23$ between the two populations, with DB1 above and DB2 below the line. It is noteworthy that the region where DB1 lies is not well-populated in \cite{dLC2020b} and there is only one galaxy from the group with small inner bars that has a bigger inner bar than DB1.    

\cite{dLC2020b} note that the population of long inner bars does not differ from the short inner bars either in terms of the parameters of the host galaxy or the bulge properties. They hypothesize that this bi-modality could be due to different formation scenarios of the double-barred structure. Based on the location of the simulated tidally induced double-barred galaxies from TNG50 in the $a_{\mathrm{IB}}$ vs $a_{\mathrm{OB}}$ plane we propose tidal interactions as discussed here as a potential explanation of the origin of the galaxies with relative and absolute big sizes of the inner bars. In the tidal scenario, the inner bar can hypothetically grow very long until the outer bar forms through the interaction. In the nuclear stellar disc scenario, the size of the inner bar is constrained by the dynamics of the outer bar, and the details of the gas content responsible for the growth of the nuclear stellar disc. The tidal nature of double-barred systems with long inner bars is also supported by the fact that NGC 3941, present in the sample of \cite{dLC2020b} (one of the green stars in Fig.~\ref{Lbars}), does not show signatures of hosting a nuclear stellar disc, as analysed in \cite{dLC2013}.

From the theoretical point of view, the two cases of tidally induced double bars from TNG50 raise an open and interesting question about the effect of tides on single-barred galaxies. It is unlikely that DB1 and DB2 are the only barred galaxies from the sample of \cite{RosasGuevara2022} that have undergone similar tidal forcing, however, they are the only ones so far that we found to have double bars. \cite{Lokas2016} carried out simulations of discs having their bars induced by a cluster-like potential and the following pericenters were inducing mostly grand-design spirals that would wind up and dissolve with time as discussed by \cite{Semczuk2017}. Although a configuration similar to a double bar was identified in these simulations, for the galaxy on the tightest orbit around the cluster, it was short-lived, in contrast to stable structures found here. In our future work, we plan to study what conditions are essential for tides to induce secondary bars. For now, it is not clear whether it is the strength of the interaction, a special orientation between the inner bar and the perturber, or the properties of the host, like gas fraction or the kinematical state of the disc.    

In this paper, we propose a new scenario for the formation of double bars in galaxies through tidal interactions. We demonstrate its viability with two example cases from the TNG50 simulations \citep{Pillepich2019}. In the proposed scenario the inner bar comes first, contrary to previous scenarios discussed in the literature, and the outer bar forms through tidal interactions. The double bars formed that way are long-lived (>3 Gyr) which is consistent with the observational findings of \cite{MA2019} suggesting the longevity of the double bars. Tidal double bars also rotate with different pattern speeds, which is consistent with simulated double bars formed in other ways. The double bars discussed here also have a very similar kinematical structure to the previously studied double bars (especially \citealt{Du2016}). The double bars formed through tidal interactions are able to reproduce large inner bars similar to those recently reported in observations by \cite{dLC2020b}.   

\section*{Acknowledgements}

 We are grateful to the IllustrisTNG team for making their simulations publicly available, and to the organizers of the "Galactic bars: driving and decoding galaxy evolution" conference held in Granada in 2023, which led to the work on this paper. We appreciate insightful discussions with T. Antoja, W. Dehnen, R. Sch{\"o}nrich, and Gaia UB team. This work was partially supported by the Spanish MICIN/AEI/10.13039/501100011033 and by "ERDF A way of making Europe" by the “European Union” and the European Union «Next Generation EU»/PRTR, through grants PID2021-125451NA-I00 and CNS2022-135232, and the Institute of Cosmos Sciences University of Barcelona (ICCUB, Unidad de Excelencia ’Mar\'{\i}a de Maeztu’) through grant CEX2019-000918-M. AdLC acknowledges financial support from the Spanish Ministry of Science and Innovation (MICINN) to the coBEARD project (PID2021-128131NB-I00) and through the Spanish State Research Agency, under the Severo Ochoa Centres of Excellence Programme 2020-2023 (CEX2019-000920-S). EA thanks the CNES for its financial support. This work was performed using the DiRAC Data Intensive service at Leicester, operated by the University of Leicester IT Services, which forms part of the STFC DiRAC HPC Facility (\url{www.dirac.ac.uk}). The equipment was funded by BEIS capital funding via STFC capital grants ST/K000373/1 and ST/R002363/1 and STFC DiRAC Operations grant ST/R001014/1. DiRAC is part of the National e-Infrastructure.

\section*{Data Availability}
No data were generated in this study.



\bibliographystyle{mnras}
\bibliography{example} 





\bsp	
\label{lastpage}
\end{document}